\documentclass[letterpaper, 10 pt, conference]{ieeeconf}  

\usepackage{balance}
\usepackage{epsf,color}
\usepackage{graphicx}
\usepackage{amsmath,bm}
\usepackage{amssymb}
\usepackage{epsfig,latexsym,amsmath,epsf,amssymb,amsfonts}
\usepackage{verbatim}
\usepackage{placeins}
\usepackage[hang]{subfigure}
\usepackage{url}
\usepackage{algorithm}
\usepackage{algorithmic}
\usepackage{caption}
\usepackage{cite}

\usepackage{soul,xcolor}
\setstcolor{red}

\newcommand{\vect}[1]{\boldsymbol{\mathbf{#1}}}

\DeclareMathOperator{\Exp}{\mathbb{E}}

\newcommand{\size}[1]{\left | #1 \right | }
\newcommand{\beam}[1]{\mathcal B_{#1 } }
\newcommand{\beambs}[1]{\mathcal B_{{\mathrm t} ,#1}}
\newcommand{\beamue}[1]{\mathcal B_{{\mathrm r},#1} }
\newcommand{\ue}{\text{UE}}
\IEEEoverridecommandlockouts                              
                                                          
\title{\LARGE \bf
Coded Energy-Efficient Beam-Alignment for Millimeter-Wave Networks
}

\author{Muddassar Hussain, Nicol\`{o} Michelusi
\thanks{This research has been funded by NSF under grant CNS-1642982.}
\thanks{Authors are with the School of Electrical and Computer Engineering, Purdue University. email: \{hussai13,michelus\}@purdue.edu.}
}

\begin{document}


\maketitle
\thispagestyle{empty}
\pagestyle{empty}
\begin{abstract}
Millimeter-wave communications rely on narrow-beam transmissions to cope with the strong signal attenuation at these frequencies, thus demanding precise alignment between transmitter and receiver. However, the beam-alignment procedure may entail a huge overhead
and its performance may be degraded by detection errors. This paper proposes a coded energy-efficient beam-alignment scheme, robust against detection errors. Specifically, the beam-alignment sequence is designed such that the error-free feedback sequences are generated from a  codebook with the desired error correction capabilities. Therefore, in the presence of detection errors, the error-free feedback sequences can be recovered with high probability. 
The assignment of beams to codewords is designed to optimize energy efficiency, and a water-filling solution is proved.
The numerical results with analog beams depict up to 4dB and 8dB gains over exhaustive and uncoded beam-alignment schemes, respectively.
\end{abstract}
\section{Introduction}
Millimeter-wave (mm-wave) technology is emerging as a promising solution to enable multi-Gbps communication, thanks to abundant bandwidth availability \cite{channel_model}. 
However, signal propagation at these frequencies poses several challenges to the design of future communication systems supporting high throughput and mobility, due to high isotropic path loss and sensitivity to blockages \cite{rappaport_book}.
To compensate the propagation loss \cite{channel_model}, mm-wave systems will leverage narrow-beam communications, by using large antenna arrays at both base stations (BSs) and user-ends (UEs).

However, narrow beams are susceptible to frequent loss of alignment due to mobility or blockage, which necessitate the use of beam-alignment protocols.
These use significant communication resources, and it is therefore imperative to 
design schemes to mitigate the overhead.
Beam-alignment in mm-wave has been a subject of intensive research.  The simplest and yet most popular scheme is \emph{exhaustive} search \cite{exhaustive}, which sequentially scans through all possible BS-UE beam pairs and selects the one with maximum signal power. A version of this scheme has been adopted in existing mm-wave standards including IEEE 802.15.3c  \cite{ieee80215c} and IEEE 802.11ad \cite{ieee80211ad}.
An interactive version has been proposed in \cite{interactiveexhaustive}, wherein the beam-alignment phase is terminated once the power of  the received beacon is above a certain threshold. 
 The second popular scheme is \emph{iterative} search \cite{iterative}, where scanning is first performed using wider beams followed by refinement using narrow beams.
 In \cite{caire}, a multiuser beam-alignment scheme  is proposed that performs beam-alignment by exploring the angle of departure (AoD) and angle of arrival (AoA)  domain through pseudorandom multi-finger beam patterns. The authors use tools from compressed sensing to estimate the AoD/AoA pair.

   In the aforementioned papers, the optimality of the corresponding search scheme is not established and the energy cost of beam-alignment is neglected, which may be significant when targeting high detection accuracy. 
    To address it, in our previous works \cite{asilomar2017, icc2018, asilomar2018},  we designed optimal energy-efficient beam-alignment protocols. 
    In \cite{asilomar2018}, we prove the optimality of a \emph{fractional search} method.
 In \cite{icc2018}, we account for the UE mobility by widening the BS beam to mitigate the uncertainty on the UE position, and optimize the trade-off between data communication and the cost  of beam-sweeping. The algorithms \cite{ita2017,asilomar2017, icc2018, asilomar2018} are designed based on the assumption that no detection errors occur in the beam-alignment phase. However, the performance may deteriorate due to mis-detection and false-alarm errors, 
  causing a loss of alignment during the communication phase. Therefore, it is of great interest to design beam-alignment algorithms robust to detection errors and, at the same time, energy-efficient.
 
Motivated by these observations, in this paper we consider the design of an energy-efficient beam-alignment protocol robust to detection errors. To do so, we restrict the solution space for the beams such that the error-free feedback sequence can only be generated from a codebook with error correction capabilities. Thus, if detection errors occur, the error-free feedback sequence may still be recovered with high probability by leveraging the structure of the error correction code. We pose the beam-alignment problem as a convex optimization problem to minimize the average power consumption, and provide its closed-form solution that resembles a "water-filling" over the beamwidths of the beam-alignment beam patterns.
   The numerical results depict the superior performance of the proposed coded technique, with up to $4$dB and $8$dB gains
 over exhaustive and uncoded beam-alignment schemes, respectively.
Open- and closed-loop error control sounding schemes have been studied in \cite{suresh}, but with no consideration on energy-efficient design.
    \emph{To the best of our knowledge, this paper is the first to propose a coded beam-alignment scheme, which is both energy-efficient and robust to detection errors.} 

In \cite{yahia}, beam-alignment is treated as a beam discovery problem in which locating beams with strong path reflectors is analogous to locating errors in a linear block code. 
Unlike \cite{yahia}, we use error correction to correct errors during the beam-alignment procedure, rather than to detect strong signal clusters.
Unlike
 \cite{ita2017, asilomar2017, asilomar2018} which rely on continuous feedback from UEs to BS, we consider a scheme where the feedback is generated only at the end of the beam-alignment phase, which scales well to multiple users scenarios.

The rest of the paper is organized as follows.
In Sec. \ref{sysmo}, we present the system model;
in Sec. \ref{optproblem}, we present the optimization problem and analysis;
in Sec. \ref{numres}, we present numerical results, followed by concluding remarks in 
Sec.~\ref{conclu}.
  
\section{system model}
\label{sysmo}
We consider a mm-wave cellular network with a single base-station (BS) and $M$ users (UEs) denoted as $\ue_i, i=1,2,\ldots,M$, in a downlink scenario. $\ue_i$ is at distance $d_i\leq  d_{\max}$ from BS,
 where $d_{\max}>0$ is the coverage radius of the BS.
We assume that there is a single strongest path between the BS and each $\ue_i$, whose angle of departure (AoD) and angle of arrival (AoA) are denoted by $\theta_{\mathrm t,i}{\sim}\mathcal U[-\pi/2,\pi/2]$ and $\theta_{\mathrm r,i}{\sim}\mathcal U[-\pi/2,\pi/2]$, respectively. $\mathcal U[a,b]$ denotes the uniform distribution over the interval $[a,b]$. We use the \emph{sectored antenna} model to approximate the beam patterns of {the} BS and UEs  \cite{sectored_model}. Under {such model}, the beamforming gain
 is characterized by the angular support of the BS and UE beams, denoted as 
$\beambs{k}{\subseteq}[-\pi/2,\pi/2]$ and $\beamue{k}{\subseteq}[-\pi/2,\pi/2]$, respectively, and is given by
\begin{align}
\label{eq:gain}
G(\beam{k}, \vect \theta_i) = \frac{\pi^2}{\size{\beam{k}}} \chi(\vect\theta_i \in \beam{k}),
\end{align}
where $\beam{k} \equiv \beambs{k} \times \beamue{k}$ and
 $\vect \theta_i \triangleq (\theta_{\mathrm t,i },\theta_{\mathrm r,i })$; $\chi(\vect\theta \in \mathcal A)$ is {the} indicator function of the set  $
\mathcal A$, and $\size{\mathcal A} \triangleq \int_{\mathcal A} \mathrm d \vect\theta$ is its Lebesgue measure.
In other words, if the AoD/AoA $\vect\theta$ lies in the beam support $\beam{k}$ of the BS and UE, then the signal is received with gain $\frac{\pi^2}{\size{\beam{k}}}$; otherwise, only noise is received.
 The received signal at $\ue_i$ can thus be expressed as 
\begin{align}
\label{sigmodel}
\vect y_k^{(i)} = h_k^{(i)}\sqrt{P_k G(\beam{k},\vect\theta_i)} \vect s_{k} + \vect n_k^{(i)},
\end{align}
where $k$ is the slot index, $\vect s_{k}$ is the transmitted sequence, $P_k$ is the transmission power of the BS, $h_k^{(i)}$ is the complex channel gain between the BS and $\ue_i$, and $\vect n_{k,i}\sim \mathcal{CN}(\vect 0, N_0 W_{\mathrm{tot}} \vect I)$ is complex additive white Gaussian noise (AWGN). The quantity $ N_0$ denotes the one-sided power spectral density of the AWGN channel  and $ W_{\mathrm{tot}}$ is the system bandwidth. We assume Rayleigh fading channels $h_k^{(i)}\sim \mathcal{CN}(0,1/\ell(d_i)), \forall i, k$, independent across UEs and i.i.d over slots, where $\ell(d_i)$ is the path loss between the BS and $\ue_i$.

\begin{figure}[!t]
\centering
  \includegraphics[width=.8\columnwidth]{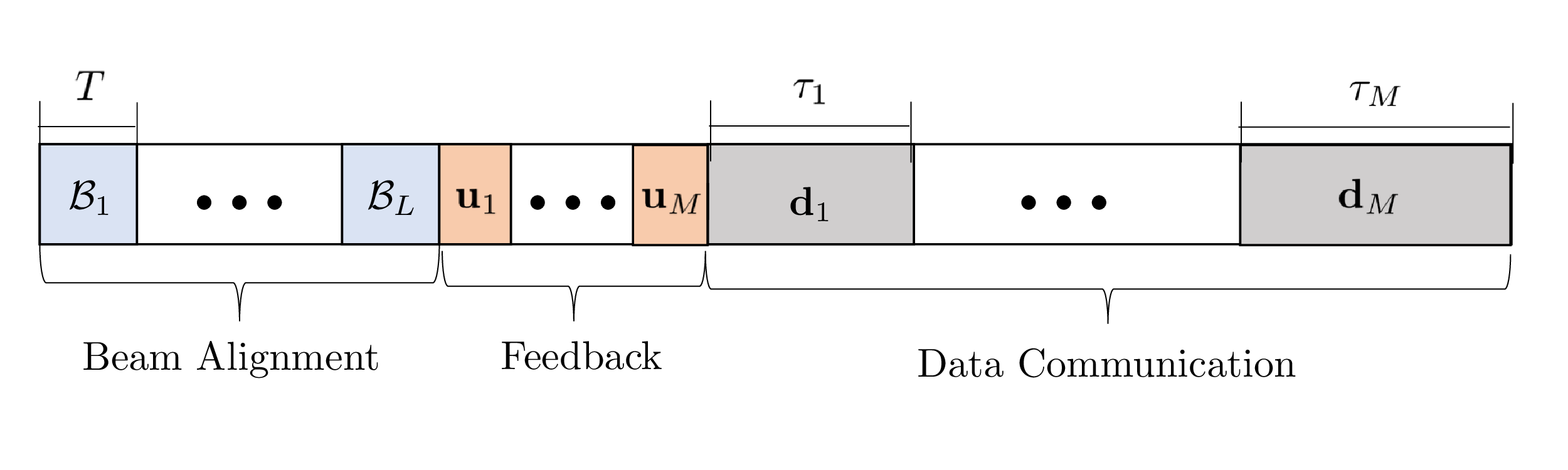}
\caption{Timing Diagram.}
\label{fig:timing_diagram}
\end{figure}

  We consider a time-slotted system where {the} frame duration $T_{\mathrm{fr}}[s]$ is divided into three phases: beam-alignment, feedback and data communication,
  of duration $T_s$, $T_{\mathrm{fb}}$
  and $T_d$, respectively, 
  with $T_s{+}T_{\mathrm{fb}}{+}T_d{=}T_{\mathrm{fr}}$,
  as depicted in Fig.~\ref{fig:timing_diagram}. Data transmission is orthogonalized across users according to a TDMA strategy. We now describe these phases in more detail.

{\bf\underline{Beam-Alignment Protocol}:}
 The beam-alignment phase, of duration $T_s$, is  divided into $L$ slots, each of duration $T = T_s/L$,  indexed by {the} set $\mathcal I_s = \{1,\ldots,L\}$. 
 In each beam-alignment slot, the BS sends a pilot sequence $\vect s_k$ using the sequence of beams $\{\beambs{k},k{=}1,\ldots,L\}$. Simultaneously, each UE receives using  the  sequence of beams $\{\beamue{k} , k{=}1,\ldots,L \}$. 
In each beam-alignment slot, $\ue_i$ tests whether $\vect\theta_i{\in}\beam{k}$ (alignment) or $\vect\theta_i{\notin}\beam{k}$ (mis-alignment). This can be expressed as the following hypothesis testing problem:
\begin{align}
\label{Hptesting}
 &\mathcal H_1:\ \vect y_k^{(i)} = h_k^{(i)}\sqrt{\frac{\pi^2 P_k}{\size{\beam{k}}}} \vect s_k + \vect n_k^{(i)},\ (\text{alignment}),\\ 
 &\mathcal H_0:\ \vect y_k^{(i)} = \vect n_k^{(i)},\ (\text{mis-alignment}).
 \nonumber
 \end{align} 
Under no CSI ($h_k^{(i)}$ unknown),
the optimal Neyman-Pearson detector for the above binary problem is the threshold detector
  \begin{align}
   \frac{|\vect s_k^H \vect y_k^{(i)}|^2}{N_0 W_{\mathrm{tot}}\| \vect s_k\|_2^2} \mathop{\lessgtr}_{\mathcal H_1}^{\mathcal H_0} \tau_{\mathrm{th}}.
   \end{align} 
   If $\ue_i$ infers that $\mathcal H_1$ is true, then it generates $u_k^{(i)}{=}1$, otherwise  $u_k^{(i)}{=}0$. 
 Each UE generates its detection sequence $\vect u_i \triangleq (u_1^{(i)},u_2^{(i)},\ldots,u_L^{(i)})\in\{0,1\}^L$ with the above detector. This is used to infer the AoD/AoA $\vect\theta_i$, and to design the beams for the data communication phase, as detailed below.
 \begin{align}
 \text{Let }\vect c_i \triangleq ( c_1^{(i)}, c_2^{(i)},\ldots, c_L^{(i)})\text{ with }c_k^{(i)} = \chi(\vect\theta_i\in \beam{k})
 \end{align}
 denote the error-free detection sequence.
 The detected $\vect u_i$ may incur mis-detection ($u_k^{(i)}{=}0$ but $c_k^{(i)}{=}1$) or false-alarm errors ($u_k^{(i)}{=}1$ but $c_k^{(i)}{=} 0$) , with probabilities
 (these can be obtained from the signal model \eqref{Hptesting})
\begin{align}
\label{pmd}
&\!\!p_{\mathrm{md},i}{=}1{-}\exp\left(-\frac{\tau_{\mathrm{th}}\size{\beam{k}}N_0 W_{\mathrm{tot}}\ell(d_i)}{\size{\beam{k}}N_0 W_{\mathrm{tot}}\ell(d_i)+P_k\pi^2\| \vect s_k\|_2^2}
\right),
\\
&\!\!p_{\mathrm{fa},i} = \exp \left( -\tau_{\mathrm{th}}  \right).
\end{align} 
 The BS transmission power $P_k$ and detector threshold $\tau_{\mathrm{th}}$ are designed to guarantee maximum error probabilities $p_{\mathrm{md},i}, p_{\mathrm{fa},i}\leq p_e$ across users (this can be achieved via appropriate beam design, see \cite{mmnets17}), which yields
\begin{align}
&\tau_{\mathrm{th}} = -\ln(p_e),\\ 
&P_k{\geq}\frac{N_0 W_{\mathrm{tot}} \ell(d_i)  }{\pi^2\| \vect s_k\|_2^2} \left[\frac{\ln(p_e)}{\ln(1-p_e)}{-}1 \right] |\mathcal B_k|,\ \forall i{\in}\{1,\ldots,M\}. 
\nonumber
\end{align}
 Equivalently, we can express the energy $E_k{\triangleq}T_{\mathrm{sy}} P_k \|\vect s_k\|^2$ as
 \begin{align}
 \label{eq:energy_ba}
 E_k \geq \phi_s \size{\beam{k}} 
\end{align}  
 where $T_{\mathrm{sy}}$ is the symbol duration; $\phi_s$ is the energy/rad$^2$ to guarantee the required detection performance among all UEs,
 \begin{align}
 \label{phis}
 \phi_s \triangleq \frac{N_0 W_{\mathrm{tot}} T_{\mathrm{sy}} }{\pi^2} \left[\frac{\ln(p_e)}{\ln(1-p_e)}-1 \right] \cdot \ell(d_{\max}).
 \end{align}
 In the rest of the paper, we enforce equality in \eqref{eq:energy_ba} for the purpose of energy-efficient beam-alignment design,
 and assume that $p_{\mathrm{md},i}=p_{\mathrm{fa},i}=p_e,\forall i$. Note that this is the worst-case scenario; in fact, in practice, an UE closer to the BS may experience a lower mis-detection probability $p_{\mathrm{md},i}<p_e$ as a result of $\ell(d_i)<\ell(d_{\max})$, see \eqref{pmd}.
 
With this notation, we write the detection sequence as
 \begin{align} 
 \vect u_i \triangleq \vect c_i \oplus \vect e_i,
 \end{align}
 where $\oplus$ denotes entry-wise modulo 2 addition, and $\vect e_i \in \{0,1\}^L$ is the beam-alignment error sequence of $\ue_i$.
 Due to the i.i.d. Rayleigh fading assumption and to the fact that false-alarm and misdetection errors occur with probability $p_{\mathrm{md},i}=p_{\mathrm{fa},i}=p_e$, independently across slots, it follows that $\vect e_i$ is independent of $\vect c_i$, and that errors are i.i.d. across UEs and slots,
with probability mass function (pmf)
\begin{align}
\label{errmodel}
 p(\vect e_i) = p_e^{W(\vect e_i)}(1-p_e)^{L-W(\vect e_i)},
 \end{align} 
 where $W(d){\triangleq}\sum_{k=1}^Ld_k$ is the Hamming weight of $\vect d{\in}\{0,1\}^L$.

 We now design a coded beam-alignment strategy, robust to detection errors. 
 If $\ue_i$ was provided  with the error-free detection sequence $ \vect c_i$, it could infer the support of $\vect \theta_i$ relative to {the} beam sequence $\{\beam{k}, k{=}1,\ldots,L\}$ to be
\begin{align}
\label{Uc}
\vect\theta_i\in \mathcal U_{\vect c_i} \triangleq \cap_{k=1}^L \beam{k}^{c_k^{(i)}},
 \end{align} 
   where we have defined
  \begin{align}
  \label{setdef}
\beam{k}^{c}=
\begin{cases}
\beam{k} & c=1,\\
[-\frac{\pi}{2},\frac{\pi}{2}]^2\setminus\beam{k} & c=0.
\end{cases}
  \end{align}
  In fact, $c_k^{(i)}{=}1\Leftrightarrow\vect\theta_i\in\beam{k}$ and 
  $c_k^{(i)}{=}0\Leftrightarrow\vect\theta_i\in[-\frac{\pi}{2},\frac{\pi}{2}]^2\setminus\beam{k}$, yielding \eqref{Uc} when considering the entire sequence $\vect c_i$. 
We let $\mathcal C$ be the set of all possible error-free detection sequences with non-empty beam support, i.e.
 \begin{align}
 \label{C}
 &\mathcal C\triangleq  \{\vect c \in \{0,1\}^L: \mathcal U_{\vect c} \neq \emptyset \},
 \end{align} 
 and $\mathcal G$ be the corresponding beam-support,
 \begin{align}
 &\mathcal G \triangleq \{\mathcal U_{\vect c}: \vect c \in \mathcal C\}. 
 \end{align} 
   Note that $(\mathcal C,\mathcal G)$ are uniquely defined by
 the beam sequence $\{\beam{k}, k{=}1,\ldots,L\}$. Likewise,
 $\{\beam{k}, k{=}1,\ldots,L\}$ is uniquely defined by
 a specific choice of  $(\mathcal C,\mathcal G)$, as can be seen by letting
\begin{align}
\label{beamk}
\beam{k} \equiv \cup_{\vect c\in \mathcal C: c_k = 1} \mathcal U_{\vect c},\ \mathcal U_{\vect c} \in \mathcal G. 
 \end{align} 
 Therefore, the problem of finding the optimal beam sequence, $\{\beam{k}, k=1,\ldots,L\}$ is equivalent to that of finding the sets $\mathcal C$ and $\mathcal G$. However, a joint optimization over $\mathcal C$ and $\mathcal G$ is intractable due to the combinatorial nature of the problem and lack of convexity. Therefore, we resort to selecting $\mathcal C$ and $\mathcal G$ independently, where $\mathcal C$ is chosen from a predefined codebook with the desired error correction capability and $\mathcal G$ is designed to optimize energy efficiency.
 
  {\bf\underline{Error Correction and Scheduling }:} 
 One way to choose $\mathcal C$ would be as all possible binary sequences of length $L$, $\mathcal C{\equiv}\{0,1\}^L$.
 However, a single error during the beam-alignment phase would result in an incorrect selection of the communication beam. For instance, in the case $L{=}3$, if the error-free codeword is $\vect c_i{=}[1,1,1]$ (and thus $\vect\theta_i{\in}\mathcal U_{[1,1,1]}$) but
 $\ue_i$ detects $\vect u_i{=}[1,0,1]$, then it will incorrectly infer that $\vect\theta_i{\in} \mathcal U_{[1,0,1]}$, resulting in outage in the data communication phase.
 
 In order to compensate for detection errors, we endow $\mathcal C$ with error correction capabilities up to $\varepsilon$ errors,
 e.g., using Hamming codes.
  Therefore, at the end of {the} beam-alignment phase, each UE applies the decoding function $f:\{0,1\}^L{\to}\mathcal C$ to {the} detection sequence $\vect u_i$. In this paper, 
  we use the minimum Hamming distance criterion to design $f(\cdot)$, i.e.,
 \begin{align}
  f(\vect u) \triangleq \arg\min_{\vect c \in \mathcal C} \| \vect u - \vect c\|_2^2.
 \end{align}
After decoding, each UE feeds back to the BS the ID of its corrected sequence $ \hat {\vect c}_i \triangleq f(\vect u_i),\ \forall i \in\{1,2,\ldots,M \}$.
{We assume that the feedback signals are received without errors at the BS, which thus}
 infers that
\begin{align}
&\vect\theta_i \in \mathcal U_{f(\vect u_i)},
\end{align}
{where $\mathcal U_{\vect d}$ is defined in \eqref{Uc}.}
Given $f(\vect u_i)$, the BS allocates the communication resources $(\tau_i, \mathcal B_{\mathrm d,i},P_i,R_i)$ to $\ue_i$ during the data communication phase,
denoting the allocated time, BS transmission power and rate, and communication beam.
In this paper, we assume a TDMA strategy, i.e., $\tau_i = T_d/M$, $\forall i\in \{ 1,2,\ldots,M\}$. 
The beam pair $\mathcal B_{\mathrm d,i}\equiv \mathcal B_{{\mathrm{t}},i}^{\mathrm d } \times  \mathcal B_{{\mathrm{r}},i}^{\mathrm d}$
is chosen as
\begin{align}
\label{eq:comm_beam}
\mathcal B_{\mathrm d,i}\equiv\mathcal B_{\mathrm d}(\vect u_i) = \mathcal U_{f(\vect u_i)}.
\end{align}
Note that, due to the error correction capability endowed in the design of $\mathcal C$, if less than (or equal to) $\varepsilon$ errors have been introduced in the beam-alignment phase, then $f(\vect u_i)=\vect c_i$, and thus correct alignment is achieved in the data communication phase ($\mathcal B_{\mathrm d,i}\equiv\mathcal U_{\vect c_i}$); otherwise, if $f(\vect u_i)\neq\vect c_i$, then the data communication beam is not aligned with the AoD/AoA, and outage occurs ($\mathcal B_{\mathrm d,i}\cap\mathcal U_{\vect c_i}\equiv\emptyset$). The resulting mis-alignment probability of $\ue_i$ can then be bounded as
\begin{align}
\label{pma}
&\!\!\!\!p_{\mathrm{ma},i}(\mathcal B_{\mathrm d}){\leq}\mathbb P(W(\vect e_i)>\varepsilon)
{=}\!\!\!\!\sum_{l=\varepsilon+1}^{L}\left(\begin{array}{c}L\\l\end{array}\right)p_e^{l}(1{-}p_e)^{L-l},\!\!
\end{align}
as per the error model \eqref{errmodel}.
Note that this is a function of $\phi_s$ via \eqref{phis},
duration $L$ of beam-alignment and number of correctable errors 
$\varepsilon$ (i.e., choice of the error correction codebook $\mathcal C$).
However, it is independent of the beam-alignment sequence $\beam{k},k{\in}\mathcal I_s$.
Therefore, the optimization over 
$\phi_s$, $L$, $\mathcal C$ and 
$\beam{k},k{\in}\mathcal I_s$ can be decoupled:
$\phi_s$, $L$, $\mathcal C$
 can be chosen to achieve a target mis-alignment performance $p_{\mathrm{ma},i}\leq p_{\mathrm{ma}}^{\max},\forall i$, whereas $\beam{k},k{\in}\mathcal I_s$ is optimized to achieve energy-efficient design. This optimization is developed in the next section.

{\bf\underline{Data Communication}:}
In the data communication phase, the BS transmits to $\ue_i$ in the assigned TDMA slot using power $P_i$ and rate $R_i$. These are designed to satisfy a maximum outage probability $p_{\mathrm{out}}(P_i,R_i) \leq \rho$, with no CSI at the transmitter ($h_k^{(i)}$ unknown at BS),
and a minimum rate constraint $R_{\min,i}$ of $\ue_i$ over the frame.
In case of mis-alignment, data communication is in outage, see \eqref{pma}.
We now consider the case of alignment, i.e., $f(\vect u_i){=}\vect c_i$ and $\vect \theta_i{\in} \mathcal B_{\mathrm{d}}(\vect u_i)$. In this case, the instantaneous signal-to-noise ratio (SNR) during the data communication slots associated with $\ue_i$ is
\begin{align}
\mathsf{SNR}_{k}^{(i)} = \frac{\pi^2\gamma_k^{(i)} P_i }{N_0 W_{\mathrm{tot}} |\mathcal B_{\mathrm{d}}(\vect u_i)|},
\end{align}
where $\gamma_k^{(i)}{\triangleq}|h_k^{(i)}|^2$. The outage probability is then given by
\begin{align*}
&p_{\mathrm{out}}(P_i,R_i)= \mathbb{P}(W_{\mathrm{tot}}\log_2(1+\mathsf{SNR}_{k}^{(i)})\leq R_i|\vect u_i)\\
&=1 - \exp\left(- (2^{\frac{R_i}{W_{\mathrm{tot}}}}-1)
\frac{\ell(d_i) N_0 W_{\mathrm{tot}}}{P_i \pi^2}|B_{\mathrm{d}}(\vect u_i) |
 \right).
\end{align*}
To meet the minimum rate constraint of $\ue_i$ over the frame, we enforce $R_i=\frac{T_{\mathrm{fr}}}{\tau_i} R_{\min,i}$.
To enforce $p_{\mathrm{out}}(P_i,R_i) \leq \rho$,\footnote{Note that the overall outage probability including mis-alignment is given by
$p_{\mathrm{ma}}^{\max}+(1-p_{\mathrm{ma}}^{\max})\rho$.} we find the power $P_i$
and the energy $E_i\triangleq P_i\tau_i$ as
\begin{align}
\label{eq:comm_energy}
E_i=\phi_{\mathrm d,i} |\mathcal B_{\mathrm{d}}(\vect u_i) |,
\end{align}
where $\phi_{\mathrm d,i}$ is the minimum energy/rad$^2$ required to meet the rate requirement of $\ue_i$ with outage probability $\rho$, given by
\begin{align}
\phi_{\mathrm d,i} \triangleq \frac{\tau_i\ell(d_i)N_0 W_{\mathrm{tot}}  \left[2^{\frac{T_{\mathrm{fr}} R_{\min,i}}{\tau_i  W_{\mathrm{tot}}}}-1\right]}{\pi^2 \ln (1/(1-\rho))}.
\end{align}

\section{Optimization Problem}
\label{optproblem}
The optimum beam-alignment design seeks to minimize the average power consumption
$\bar P_{\mathrm{avg}}(\mathcal B)$
 of the BS, over the beam-sequence $\mathcal B{=}\{\beam{k},k\in\mathcal I_s\}$ in the beam-alignment phase, i.e.,
\begin{align}
\label{OP}
\vect{P1}:\  \mathcal B^*=\arg\min_{\mathcal B}\bar P_{\mathrm{avg}}(\mathcal B),
\end{align}
where, using \eqref{eq:energy_ba} and \eqref{eq:comm_energy}, 
$\bar P_{\mathrm{avg}}(\mathcal B)$ is given by
\begin{align}
\label{eq:avg_power}
\bar P_{\mathrm{avg}}(\mathcal B) 
=
 \frac{1}{T_{\mathrm{fr}}}\Exp \left [\sum_{k=1}^{L} \phi_s \size{\beam{k}} +
 \sum_{i=1}^M \phi_{d,i} \size{\mathcal B_{\mathrm d}(\vect u_i) }  \right],
\end{align}
with  $\mathcal B_{\mathrm d}(\vect u_i)$ given by \eqref{eq:comm_beam}. The expectation
 is over the detected and error-free sequences $\{(\vect u_i,\vect c_i),i=1,\ldots,M\}$.

Using \eqref{beamk} and \eqref{eq:comm_beam}, we can express the beam-alignment and data communication beams as
\begin{align}
\label{eq:transf}
\mathcal B_k=\cup_{\vect d \in \mathcal C:d_k=1}\mathcal U_{\vect d},
\quad\mathcal B_{\mathrm d}(\vect u_i) = \mathcal U_{f(\vect u_i)}.
\end{align}
In fact, $\mathcal U_{\hat{\vect c}_i}$ represents the {estimated} support of the AoD/AoA of $\ue_i$, when it detects the error corrected sequence $\hat{\vect c}_i{=}f(\vect u_i)$. Note that $\{\mathcal U_{\vect d}:\vect d{\in}\mathcal C\}$ forms a partition of the 
AoD/AoA space $[-\pi/2,\pi/2]^2$. In fact, using \eqref{Uc}, the fact that $\cap_{k=1}^L\mathcal B_k^{d_k}\equiv\emptyset,\forall\vect d\notin\mathcal C$,
and the set definition \eqref{setdef},
 we can show that
   \begin{align}
&\cup_{\vect d\in\mathcal C}
  \mathcal U_{\vect d}\equiv
  \cup_{\vect d\in\{0,1\}^L}
  \cap_{k=1}^L\mathcal B_k^{d_k}=[-\pi/2,\pi/2]^2,
\\&
   \label{intersect}
  \mathcal U_{\vect d_1}\cap\mathcal U_{\vect d_2}
  \equiv
  \cap_{k=1}^L[\mathcal B_k^{d_{1,k}}\cap\mathcal B_k^{d_{2,k}}]
  \equiv\emptyset,\quad \forall \vect d_1\neq\vect d_2.
  \end{align}
Therefore,
letting $\omega_{\vect d} \triangleq \size{\mathcal U_{\vect d}}$
be the beamwidth of the sector $\mathcal U_{\vect d}$ and using \eqref{eq:transf},
 we can rewrite the average power as
\begin{align*}
&\bar P_{\mathrm{avg}}(\omega){=}
 \frac{1}{T_{\mathrm{fr}}}\Exp \Biggr[\sum_{k=1}^{L} \phi_s \size{\cup_{\vect d \in \mathcal C:d_k=1}\mathcal U_{\vect d} }\\&{+}
 \sum_{i=1}^M \phi_{d,i}\Bigr\{ \size{\mathcal U_{\vect c_i }} \chi(W(\vect e_i){\leq}\varepsilon)
 + \size{\mathcal U_{f(\vect c_i \oplus \vect e_i )}} \chi(W(\vect e_i){>}\varepsilon) \Bigr\}\Biggr]\\
&\stackrel{(a)}{=}   \frac{1}{T_{\mathrm{fr}}}\Exp \Biggr[\sum_{k=1}^{L} \phi_s \sum_{\vect d \in \mathcal C: d_k=1}\omega_{\vect d}
\\&+
 \sum_{i=1}^M \phi_{d,i}\Bigr\{ \omega_{\vect c_i } \chi(W(\vect e_i)\leq \varepsilon)
 + \omega_{f(\vect c_i \oplus \vect e_i )} \chi(W(\vect e_i)> \varepsilon) \Bigr\}   \Biggr],
 \end{align*}
 where  in (a) we used the facts that $\{\mathcal U_{\vect c}{:}\vect c{\in}\{0,1\}^L\}$ is a partition of $[-\pi/2,\pi/2)^2$ and that, if fewer than $\varepsilon$ errors occur in the beam-alignment phase, then the support of $\vect\theta_i$ is detected correctly.
 Note that, since the AoD/AoA pair $\vect \theta_i$  is uniformly distributed in the space $[-\pi/2,\pi/2)^2$, the probability of occurrence of the error-free sequence $\vect c_i{=}\vect x$ is
 \begin{align}
 \mathbb P(\vect c_i = \vect x) =  \mathbb P(\vect \theta_i \in \cap_{k=1}^{L}  \beam{k}^{x_k}) = \mathbb P(\vect \theta_i \in \mathcal U_{\vect x}) = \frac{\omega_{\vect x}}{\pi^2},
 \end{align}
 while the error sequence $\vect e_i, \forall \vect e_i \in \{0 ,1 \}^L$ follows the pmf $p(\vect e_i)$ given in \eqref{errmodel}.
 This leads to
 \begin{align}
\label{eq:opt_prob2b}
&\bar P_{\mathrm{avg}}(\omega)= 
\frac{1}{T_{\mathrm{fr}}} \Biggr[\phi_s \sum_{\vect d \in \mathcal C}W(\vect d)\omega_{\vect d}  
\\&+ \frac{M \bar\phi_d}{\pi^2 }\sum_{\vect c \in \mathcal C} \Bigr\{\omega_{\vect c}^2\ \mathbb P(W(\vect e)\leq \varepsilon) +\!\!\!\!\!\!\!\!\!\!\!\!\sum_{\vect e \in \{0,1 \}^L:W(\vect e)>\epsilon}\!\!\!\!\!\!\!\!\!\omega_{f(\vect c \oplus \vect e)} \omega_{\vect c} p(\vect e)
 \Bigr\} 
 \Biggr], \nonumber
 \end{align}
 where we used the fact that
 $$\sum_{k=1}^{L} \sum_{\vect d \in \mathcal C: d_k=1}\omega_{\vect d}=
 \sum_{\vect d \in \mathcal C}\sum_{k=1}^{L}\chi(d_k=1)\omega_{\vect d} =\sum_{\vect d \in \mathcal C}W(\vect d)\omega_{\vect d},$$ and we have defined
 $\bar{\phi}_d \triangleq \frac{1}{M}\sum_{i=1}^M \phi_{d,i}$. 
 Thus, the optimization problem \vect{P1} can be restated as that of optimizing the "beamwidths" $\omega_{\vect d},\vect d\in\mathcal C$. The sequence of beams with desired beamwidth solution of this optimization problem can then be obtained via \eqref{eq:transf}, where $\size{\mathcal U_{\vect d}}=\omega_{\vect d}$.
Note that $\omega_{\vect d} \triangleq \size{\mathcal U_{\vect d}}$
needs to satisfy the constraint $\sum_{\vect d\in\mathcal C}\omega_{\vect d}=\pi^2$, 
since $\{\mathcal U_{\vect d},\vect d\in\mathcal C\}$ is a partition of $[-\pi/2,\pi/2]^2$.
 
  However, it can be shown that the cost function $\bar P_{\mathrm{avg}}(\omega) $ is non-convex with respect to $\omega$,
 due to the quadratic terms $\omega_{f(\vect c \oplus \vect e)} \omega_{\vect c}$ appearing in \eqref{eq:opt_prob2b}.
 In order to overcome this limitation, we propose to upper bound \eqref{eq:opt_prob2b} by a convex function. To determine this upper bound,
 note that the partition constraint $\sum_{\vect d\in\mathcal C} \omega_{\vect d} = \pi^2$
  and $\omega_{\vect d} \geq 0, \forall \vect d \in \mathcal C$
 imply that
 $\omega_{f(\vect c \oplus \vect e)} \leq \pi^2$.  Thus,  we upper bound \eqref{eq:opt_prob2b} as
\begin{align}
\label{eq:opt_prob3} 
&\bar P_{\mathrm{avg}}(\omega) 
\leq \frac{1}{T_{\mathrm{fr}}} \Biggr[\phi_s \sum_{\vect d \in \mathcal C}W(\vect d)\omega_{\vect d}  
\\&
 {+}
 \frac{M \bar\phi_d}{\pi^2 }\sum_{\vect c \in \mathcal C} \Bigr\{\mathbb P(W(\vect e){\leq} \varepsilon) (\omega_{\vect c}^2 -  \pi^2 \omega_{\vect c}) {+}\pi^2 \omega_{\vect c}
 \Bigr\} 
 \Biggr] \triangleq  \hat P_{\mathrm{avg}}(\omega).\nonumber
 \end{align}
Note that, if the probability of incurring more than $\varepsilon$ errors is made sufficiently small (by appropriately choosing the error correction code $\mathcal C$), say $\mathbb P(W(\vect e)> \varepsilon)\leq \delta\ll 1$, then we can bound the gap $\hat P_{\mathrm{avg}}(\omega) -\bar P_{\mathrm{avg}}(\omega) $ by
\begin{align}
0\leq\hat P_{\mathrm{avg}}(\omega) -\bar P_{\mathrm{avg}}(\omega) \leq \frac{M \bar\phi_d \pi^2}{T_{\mathrm{fr}}} \delta,
  \end{align}
  
 Thus, we consider the minimization of the upper bound $\hat P_{\mathrm{avg}}(\omega)$ instead of the original function $\bar P_{\mathrm{avg}}(\omega)$, yielding the optimization problem
   \begin{align}
\vect{P2}:\  \omega^*=\arg\min_{\omega\geq 0}\hat P_{\mathrm{avg}}(\omega)\ \text{s.t. }
  \sum_{\vect d\in \mathcal C}\omega_{\vect d}=\pi^2,
  \end{align}
 We now study the optimization problem $\vect{P2}$.
Note that this is a convex quadratic problem with respect to $\omega_{\vect c}:\vect d\in\mathcal C$.
The dual function associated with $\vect{P2}$ is given by
  \begin{align} \nonumber
&g(\mu)
=\min_{\omega\geq 0}
\hat P_{\mathrm{avg}}(\omega) -\mu\left(\sum_{\vect d\in\mathcal C}\omega_{\vect d}-\pi^2\right),
  \end{align}
  whose minimizer yields the "water-filling" solution
    \begin{align}
 \omega_{\vect d}^*
  =
\frac{\pi^2\phi_s}{2\mathbb P(W(\vect e)\leq \varepsilon)M\bar{\phi}_d} \Bigr[\lambda-W(\vect d)
  \Bigr]^+.
  \end{align}
The dual variable $\lambda$ is chosen so as to satisfy the constraint
\begin{align}
\sum_{\vect d\in\mathcal C}\omega_{\vect d}^*=\pi^2,
\end{align}
or equivalently, as the unique solver of 
\begin{align}
h(\lambda)=\frac{\phi_s}{2\mathbb P(W(\vect e)\leq \varepsilon)M\bar{\phi}_d} \sum_{w=0}^L n_w\left[\lambda-w
  \right]^+
=1,
\end{align}
where $n_w{\triangleq}\sum_{\vect c \in \mathcal C} \chi\left(W(\vect c){=}w\right)$ is {the} number of codewords in {the} codebook $\mathcal C$ with Hamming weight equal to $w$. 

The optimal dual variable $\lambda^*$ can be found using the bisection method
over the interval $[\lambda_{\min},\lambda_{\max}]$.
 In fact,
$h(\lambda)$ is a non-decreasing function of $\lambda>0$, with
$h(0)=0$ and, 
using the fact that $\left[\lambda-w
  \right]^+\leq\lambda$, we find that
  $$h(\lambda)\leq \frac{\phi_s}{2\mathbb P(W(\vect e)\leq \varepsilon)M\bar{\phi}_d} \lambda |\mathcal C|,$$
  where $|\mathcal C|$ is the cardinality of $\mathcal C$,
   hence $\lambda^*{\geq}\frac{2 M\bar {\phi}_d P(W(\bm e)\leq \varepsilon) }{|\mathcal C| \phi_s}$. Moreover, by denoting $\overline W \triangleq \frac{1}{|\mathcal C|}\sum_{w=0}^L n_w w$ as the average weight of the codewords in $ \mathcal C$, we observe that
\begin{align*}
&\sum_{w=0}^L n_w
\left[\lambda- w\right]
=
[\lambda -  \overline{W}]|\mathcal C|
\leq 
\frac{2 M\bar {\phi}_d P(W(\bm e)\leq \varepsilon) }{ \phi_s}
h(\lambda),
\end{align*}
thus implying the following upper and lower bounds to $\lambda^*$,
\begin{align*}
\lambda_{\min}\triangleq
\frac{2  M\bar{\phi}_d P(W(\bm e)\leq \varepsilon) }{|\mathcal C| \phi_s}{\leq}\lambda^*{\leq} \lambda_{\min}{+}\overline W\triangleq
\lambda_{\max}.
\end{align*}
\section{Numerical Results}
\label{numres}
In this section, we compare the performance of the proposed scheme with other schemes. We use Monte-Carlo simulation with $10^5$ iterations for each simulation point. The common simulation parameters used are as follows: $T_{\mathrm{fr}}{=}20$ms, $T{=}10\mu$s, $[\text{Number of BS antennas}]{=}64$, $[\text{Number of UE antennas}] = 1$, $[\text{BS-UE separation}] {=} 10$m, $N_0{=} -173$dBm, $W_{\mathrm{tot}}{=} 500$MHz, $[\text{carrier frequency}]{=}30$GHz, $\phi_s{=}6$dBm, and $\rho {=} 10^{-3}$. Moreover, we use the beamforming algorithm in \cite{song} to generate the beamforming codebook.
With these values, we have observed numerically that the probability of detection errors is in the range $p_e{\in}[0.1,0.3]$, due not only to noise and the Rayleigh fading channel, but also to sidelobes, which are not accounted for in the hypothesis testing problem \eqref{Hptesting}.
Thus, we set $p_e{=}0.3$ to capture this more realistic scenario.

\par
\begin{figure}[!t]
\centering
  \includegraphics[width=.8\columnwidth]{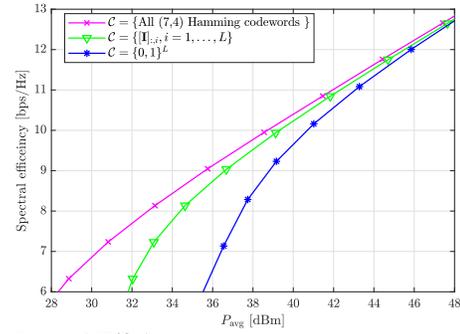}
\caption{Spectral Efficiency versus average power consumption.}
\label{fig:spec}
\end{figure}

 In Fig.~\ref{fig:spec}, we depict the spectral efficiency ($\text{Throughput}/W_{\mathrm{tot}}$) versus the average power consumption. The curves correspond to three different choices of the codebook $\mathcal C$: the  Hamming codebook $\mathcal C{=}(7,4)$, representing the proposed coded energy-efficient scheme, with error correction capability up to $\varepsilon{=}1$ errors; $\mathcal C{=}\{[\vect I]_{:,i},i{=}1,\ldots,L \}$, representing the exhaustive search scheme, where
$[\vect I]_{:,i}$ denotes the $i$th column of the $L\times L$ identity matrix $\vect I$; and $\mathcal C = \{0,1 \}^L$, representing the scheme with no error correction capabilities (uncoded).
 We use $L{=}16$ for the exhaustive search scheme, and 
 $L{=}7$ for the coded and uncoded schemes. In the figure, we observe that the proposed scheme using (7,4) Hamming codebook outperforms the other two schemes, thanks to its error correction capabilities,
 with a performance gain up to  $4$dB over exhaustive and $8$dB over the uncoded scheme. Surprisingly, the exhaustive scheme exhibits superior performance compared to the uncoded scheme, despite its more significant time overhead ($L{=}16$ vs $L{=}7$). This can be attributed to the fact that 
 the codewords in the exhaustive codebook exhibit a minimum Hamming distance of $2$, whereas the uncoded codebook exhibits minimum Hamming distance equal to $1$,
 and is thus more susceptible to detection errors during the beam-alignment phase.
\section{Conclusions}
\label{conclu}
In this paper, we have designed a coded energy-efficient beam-alignment. The scheme minimizes power consumption and uses an error correction code to recover from detection errors introduced during beam-alignment. We compare our proposed scheme with energy-efficient uncoded beam-alignment and exhaustive search, demonstrating  its superior performance.


\balance
\bibliographystyle{IEEEtran}
\bibliography{IEEEabrv,biblio} 

\end{document}